\documentclass[prd,byrevtex,preprint,
a4paper,
showkeys
]{revtex4}
\usepackage{amsmath,amssymb}
\usepackage{graphicx}
\newcommand{\bs}[1]{\boldsymbol{#1}}
\begin{document}
\title{Hoop conjecture for colliding black holes :\\
non-time-symmetric initial data\\
}
\author{Hirotaka Yoshino}
\email{hyoshino@allegro.phys.nagoya-u.ac.jp}
\author{ Yasusada Nambu}
\email{nambu@allegro.phys.nagoya-u.ac.jp}
\author{ Akira Tomimatsu}
\email{atomi@allegro.phys.nagoya-u.ac.jp}
\affiliation{Department of Physics, Graduate School of Science, Nagoya
University, Chikusa, Nagoya 464-8602, Japan}
\date{\today}
\begin{abstract}
  The hoop conjecture is well confirmed in momentarily static spaces,
  but it has not been investigated systematically for the system with
  relativistic motion. To confirm the hoop conjecture for
  non-time-symmetric initial data, we consider the initial data of
  two colliding black holes with momentum and search an apparent horizon
  that encloses two black holes.  In testing the hoop conjecture, we use
  two definitions of gravitational mass :
  one is the ADM mass and the other is the quasi-local mass defined by
  Hawking.  Although both definitions of gravitational mass
  give fairly consistent picture of the hoop
  conjecture, the hoop conjecture with the Hawking mass can judge the
  existence of an apparent horizon for wider range of parameters of the 
  initial data compared to the ADM mass.
  \end{abstract}
\maketitle

\section{introduction}
The concept of an apparent horizon is very important to understand the
global feature of spacetimes because the formation of an apparent
horizon implies the existence of an event horizon outside of it, and a black
hole has formed.  Hence it is of great interest to investigate the
condition for the apparent horizon formation.  But the necessary and
sufficient condition for the existence of an apparent horizon is not
precisely understood so far.  Although one theorem that gives a
sufficient condition for the apparent horizon formation was proved by
Schoen and Yau~\cite{SY}, it cannot be
applicable to vacuum spaces, and it does not give a necessary
condition.  As one possibility to give a necessary and sufficient
condition for the black hole formation, there is the hoop
conjecture~\cite{MWM}.  The conjecture states that for a spacelike
hypersurface $\Sigma$, a black hole horizon exists if and only if
the mass $M$ of the system gets compacted into the interior of a closed surface S
whose circumference (hoop) $C$ satisfies the condition
\begin{equation}
H\equiv \frac{C}{4\pi M}\lesssim 1.
\end{equation}
  Its physical meaning is that the mass concentration in
all direction is essential for the formation of a horizon, and the ratio
$H$ is a parameter to judge the horizon formation.

It is meaningful to confirm the hoop conjecture not only because it is
interesting theoretically, but also it can be a useful tool to test
the formation of a black hole in numerical simulations.  If we want to
know whether a black hole has formed, we must solve a differential
equation that determines the location of apparent horizons.  If the hoop 
conjecture is
correct, we can judge the black hole formation by calculating the hoop
$C$ and the mass $M$ and then taking their ratio. Therefore the hoop
conjecture has a possibility to give an easier method to judge the
black hole formation.

Several tests of the hoop conjecture has been done by various authors.
Shapiro and Teukolsky~\cite{ST} calculated the gravitational collapse
of collisionless gas spheroids numerically.  They found that when a
naked singularity appears, the ratio $H$ becomes greater than $1$ and when an
apparent horizon forms, $H$ takes a value close to $1$.  Nakamura
{\it et al.}~\cite{NST} investigated an analytic series of the initial
data for momentarily static dust spheroids.  Their result shows that
there does not exist a critical value $H$ that gives both the necessary and 
the sufficient condition for the apparent horizon formation.  Nevertheless, 
they indicated that there exist two characteristic values of $H$ ; 
the larger one gives the necessary condition and the smaller one gives 
the sufficient condition for the horizon formation. 
This means that the value of $H$ can still be a 
useful
parameter for the apparent horizon formation. 
Other tests with
momentarily static initial data also show $H$ is a parameter for the
horizon formation~\cite{AHST,CNNS,Chiba}, and they support the hoop conjecture.

In this paper, we concentrate on the test of the hoop conjecture using
initial data of Einstein's equation.
Although there are several tests
of the hoop conjecture, its validity for non-time-symmetric initial data
has not been systematically
investigated.  We pay attention to the effect of the
motion of gravitational bodies on the apparent horizon formation.
We prepare the initial data
of colliding black holes with equal mass and equal linear
momentum. We use the solution obtained by Bowen and York~\cite{BY}.
The initial data depends on two parameters : one is a
separation $L$ of two black holes and the other is a momentum $P$ of
each black hole.  In the momentarily static case, it is well known
that the decrease of the separation $L$ of two black holes causes the
formation of an apparent horizon that envelopes both black holes.  It
is this horizon that we concentrate on in this paper.  We search the
parameters $L$ and $P$ that lead to the formation of an apparent horizon.

In testing the hoop conjecture, we must specify the definition of the
circumference $C$ and the mass $M$.  As the circumference $C$ of the
hoop, Chiba \textit{et al.}~\cite{CNNS} has proposed the appropriate
 definition for all $S^2$ surfaces.
As the definition of the mass
$M$, the ADM mass is adopted usually~\cite{NST,AHST,CNNS,Chiba}.
But intuitively, the formation of an apparent horizon occurs if the mass
$M$ of the system is concentrated in a sufficiently small region.  The
ADM mass represents the total energy of the system and we expect that
$H$ given by the ADM mass becomes a good parameter to judge
the  horizon formation only if the energy of the
system is well concentrated in a region enclosed by a some compact 
surface S on which the circumference $C$ is evaluated.
When a lot of energy is distributed outside the surface S, the ADM
mass does not give a correct gravitational mass contained interior of
 the surface S
and is not suitable for testing the hoop conjecture.

Actually, if the ADM mass is used, we have some examples that $H$ is
not a good parameter for the horizon formation.
One example is a charged star~\cite{B},  for which  $H$ can take
a value close to $0.5$ even if an apparent horizon does not exist.
Flanagan~\cite{F} suggested that one should use ``quasi-local mass''
that extract gravitational energy contained  interior of a
closed surface, because the
energy of electric field distributes outside the surface of a star.
In fact,  if we use quasi-local mass, we can show that 
$H$ becomes larger than unity for such an equilibrium star.
As one cannot
uniquely determine the local energy in general relativity, there are
many definitions of quasi-local mass. We do not know what quasi-local
mass is best for the hoop conjecture.  In the case of a charged star,
the definitions of the quasi-local mass due to
Hawking~\cite{H} and Penrose~\cite{P,T} give better results than the ADM mass.
Another example is the Brill wave spacetime~\cite{AHST}. In this
case, the gravitational energy cannot be  localized so much.  
If we consider the Brill wave with large width,
$H$ given by the ADM mass cannot be a good parameter to judge the horizon
formation~\cite{AHST}.  These
two examples indicates that $H$ given by the quasi-local mass becomes a good
parameter to judge the horizon formation.

For the Bowen-York initial data, 
Jansen {\it et al.}~\cite{JDKN} discussed that this space
is not a pure black hole system. They calculated the mass and
the scalar invariants of the Bowen-York initial data
and compared them with those of the boosted Schwarzschild black hole initial data.
Their result shows that the Bowen-York initial data
for the moving black hole system contains a junk gravitational field at large
distances from black holes. Hence, we  expect that $H$ with the
quasi-local mass can be a better parameter than  $H$ with the ADM mass.

In this paper, we calculate $H$ using two different definition of the mass:
 the ADM mass and the quasi-local mass.
As the quasi-local mass, we use Hawking's quasi-local mass~\cite{H} because
for spherically symmetric spaces, $H$ given by the Hawking mass has a critical
 value $H=1$ for the horizon formation ; 
 if $H\le 1$, there exists a horizon and
if $H> 1$, there is no apparent horizon.
Thus $H$ given by the Hawking mass is a critical parameter
 for the horizon formation as far as the spherically symmetric cases are concerned.
For axially symmetric cases, we expect that this desirable feature of the
Hawking mass will hold. Even there is no critical value,
we regard $H$ as a parameter to judge  the horizon formation
if there exist two values $H_1$ and $H_2$ ; for $H<H_1$, an apparent horizon
 exists and for $H>H_2$, an apparent horizon does not exist.
  Once $H$ are obtained as 
functions of the parameters $L$ and $P$, we can define two regions in the parameter
 space $(L, P)$:
\begin{align}
 &\mathcal{R}_1\equiv\{(L,P)|~H(L, P)<H_1\},\notag \\
 &\mathcal{R}_2\equiv\{(L,P)|~H(L, P)>H_2\}.
\end{align}
We compare the regions $\mathcal{R}_{1,2}$ for the ADM mass with those 
for the Hawking mass. 
 If $\mathcal{R}_{1,2}(\rm{ADM})\subset \mathcal{R}_{1,2}(\rm{Hawking})$, 
 we can say that $H$ with the Hawking mass is 
 better than $H$ with the ADM mass  because the former predict wider range 
 of the initial data parameter on the  existence of the apparent horizon.

This paper is organized as follows. In Section II, we explain the
method of our analysis of the two black hole initial data. In Section
III, we investigate the hoop conjecture for two black holes by estimating 
the  ADM mass and the Hawking's quasi-local mass.  Section IV is devoted to summary
and discussion. We adopt the units of $c=G=1$.

\section{Initial data for two black holes}

\subsection{Initial value equations and apparent horizons}

Let $\Sigma$ be a spacelike hypersurface with the intrinsic three
metric $h_{ij}$ and the extrinsic curvature $\mathcal{K}_{ij}$.  The
initial value equations for a vacuum space are
\begin{align}
  &{}^{(3)}R-h^{im}h^{jn}\mathcal{K}_{ij}\mathcal{K}_{mn}
+\mathcal{K}^2=0,\\
  & ~D^i(\mathcal{K}_{ij}-h_{ij}\mathcal{K})=0,
\end{align}
where ${}^{(3)}R$ is the Ricci scalar of the spacelike hypersurface
$\Sigma$, $\mathcal{K}$ is the trace of the extrinsic curvature and
$D_i$ is the induced covariant derivative on $\Sigma$.  We assume that the
space is conformally flat and maximally sliced:
\begin{equation}
h_{ij}=\Psi^4 f_{ij}, \quad \mathcal{K}=0,
\end{equation}
where $f_{ij}$ is the flat space metric.  Taking
$\mathcal{K}_{ij}=\Psi^{-2} K_{ij}$, the initial value equation
becomes
\begin{align}
&\partial^2\Psi=-\frac{1}{8}K^{ij}K_{ij}\Psi^{-7},\label{conformal}\\
&\partial^i K_{ij}=0,\label{extrinsic}
\end{align}
where $\partial_i$ is the partial derivative of a flat space and
$\partial^2$ is the flat space Laplacian.  We use the solution of
 Eq.~\eqref{extrinsic} obtained by Bowen and York~\cite{BY}:
\begin{equation}
K_{ij}=\frac{3}{2r^2}[
P_i n_j +P_j n_i- \left( \delta_{ij}-n_in_j \right)
P^k n_k ]~, \label{eq:by-sol}
\end{equation}
where $n^i=x^i/r$ and $P^i$ is the linear momentum defined by
\begin{equation}
P_i=\frac{1}{8\pi}\int_{r\rightarrow\infty}
(\mathcal{K}_{ij}N^j-\mathcal{K}N_i)dS.
\end{equation}
Because Eq.\ (\ref{extrinsic}) is a linear equation, one can get a solution 
for two black holes with  momentum by superposing the solution
\eqref{eq:by-sol} as
\begin{align}
&K_{ij}=K_{ij}^{(+)}+K_{ij}^{(-)}, \\
&K_{ij}^{(\pm)}=\frac{3}{2r^{(\pm)}{}^2}[
P_{i}^{(\pm)} n_{j}^{(\pm)} +P_{j}^{(\pm)} n_{i}^{(\pm)}- \left(
\delta_{ij}-n_{i}
^{(\pm)}n_{j}^{(\pm)} \right)
P^{(\pm)k} n_{k}^{(\pm)} ] .
\end{align}
The location of each black hole is $(0,0,\pm L)$  and
\begin{equation}
P_{i}^{(\pm)}=\left(0,0,\mp P \right),\quad
n_{i}^{(\pm)}={\left(x,y,z\mp L \right)}/{r^{(\pm)}},
\end{equation}
where $r^{(\pm)}=\sqrt{x^2+y^2+(z\mp L)^2}$. For the collision of
black holes, $P$ is positive. Eq.\eqref{conformal} can be rewritten
as the form of integral equation
\begin{equation}
\Psi=1+\frac{M_0}{4r^{(+)}}+\frac{M_0}{4r^{(-)}}
+\frac{1}{4\pi}\int\frac{K^{ij}K_{ij}\Psi^{-7}}{8|\bs{r}
-\bs{r}'|}d\bs{r}'{}^3.\label{formal}
\end{equation}
For $P=0$, this expression reduces to the solution of the
Brill-Lindquist two black hole initial data that has two
Einstein-Rosen bridges~\cite{BL}. Each singular point $r^{(\pm )}=0$
corresponds to the asymptotically flat region.  Thus the solution of 
Eq.~\eqref{formal} represents a space with three asymptotically flat
regions.  We use this expression to specify the boundary condition at
$r^{(\pm )}=0$.  For sufficiently small value of $r^{(\pm)}$,
Eq.\eqref{formal} reduces to
\begin{equation}
\Psi\approx C+\frac{M_0}{4r^{(\pm)}},
\end{equation}
and the boundary condition at $r^{(\pm)}=0$ can be written
\begin{equation}
\frac{d\Psi}{dr^{(\pm)}}=-\frac{M_0}{4r^{(\pm)}{}^2}
\quad \text{as}\quad r^{(\pm)}\rightarrow 0\label{boundary}.
\end{equation}
We introduce a small cut off parameter 
$r^{(\pm)}_{\text{cut}}=L/\cosh(10)\approx 10^{-4}L$
to impose the boundary condition at $r^{(\pm)}=0$ in numerical
integration. By using a different value of cut off parameter
$r^{(\pm)}_{\text{cut}}=L/\cosh(9)$, we estimated the numerical error
due to the cut off is $0.03\%$.  The other boundary condition is
\begin{equation}
\frac{\partial\Psi}{\partial\rho}\Big|_{\rho=0}=
\frac{\partial\Psi}{\partial z}\Big|_{z=0}=0,
\end{equation}
where $\rho=\sqrt{x^2+y^2}$.  This comes from the axial symmetry about $z$-axis and
the mirror symmetry about the equatorial plane.  We solve Eq.~\eqref{conformal} numerically with these boundary conditions by a
finite difference method with $160\times 100$ grids in bispherical
coordinates.  By comparing the calculation with $320\times
200$ grids, we estimate the numerical error is about $0.2\%$.

Once the initial data are obtained, we search apparent horizons for the
initial data. The apparent horizon is a $S^2$ surface on which the expansion of
the outgoing null geodesic congruence normal to it vanishes.  Let
$s^a$ be a unit vector tangent to $\Sigma$ and normal to a two-dimensional 
spacelike surface and $n^a$ be a unit vector normal to $\Sigma$.  Then 
$k^a=n^a+s^a$ is a
future directed and outward pointing null vector on the surface.  This 
surface is an
apparent horizon if the condition
\begin{equation}
\theta_{+}=\nabla_ak^a=D_as^a-K+K_{ab}s^as^b=0\label{apparent}
\end{equation}
is satisfied.  We rewrite this equation with spherical coordinates
$(r,\theta,\phi)$.  By expressing the $S^2$ surface  as
$r=h(\theta)$,
Eq.~\eqref{apparent} becomes
\begin{multline}
  h_{,\theta\theta}-
  \left(4\frac{\Psi_{,r}}{\Psi}+
    \frac{2}{h}\right) h^{2}
  -\left(4\frac{\Psi_{,r}}{\Psi}+
    \frac{3}{h}\right)h_{,\theta}^{2}
  +\left(4\frac{\Psi_{,\theta}}{\Psi}+\cot\theta\right)
  h_{,\theta}\left(1+\frac{h_{,\theta}^{2}}{h^{2}}\right) \\
  -h^{2}\left( 1+\frac{h_{,\theta}^{2}}{h^{2}}
  \right)^{1/2}
  \frac{1}{\Psi^{4}}\left(
    K_{rr}-2\frac{h_{,\theta}}{h^2}K_{r\theta}
    +\frac{h_{,\theta}^2}{h^4}K_{\theta\theta}
  \right)=0.
\end{multline}
We solve this equation with the boundary condition $h_{,\theta}=0 ~
\text{at}~\theta=0,\pi/2$.

\subsection{Circumference, ADM mass and Hawking mass}
In axially symmetric spaces, the circumference $C$ of a surface S is defined 
by
\begin{equation}
C(\text{S})={\text{max}} (L_e,L_p),
\end{equation}
where $L_e$ is the maximum length of closed azimuthal curves and $L_p$
is the twice of the distance from the north pole to the south pole.
We survey all surfaces S which enclose two black holes.  Because
the circumference can become arbitrarily large, we must calculate the
minimum value $H$ of $H(\rm{S})$ for all surfaces S.

We write $H$ given by the ADM mass as $H^{(\rm{A})}$ hereafter. Searching
the minimum value $H^{(\rm{A})}$
of $H^{(\rm{A})}(S)$ corresponds to finding the surface
with the minimum value of $C$ because the ADM mass is defined independent of the
 surface S.
For our initial data, $L_e<L_p$ is
always satisfied and we have
\begin{equation}
  C=L_p=2\int^{\pi}_0\Psi^2\sqrt{{h_{,\theta}}^2+h^2}\,d\theta~,
\end{equation}
where we express the surface
as $r=h(\theta)$.  We obtain the equation
to determine the hoop length by taking variation of this integral $\delta 
C=0$ :
\begin{equation}
  h_{,\theta\theta}
  -\left(2\frac{\Psi_{,r}}{\Psi}+
    \frac{1}{h}\right) h^{2}
  -\left(2\frac{\Psi_{,r}}{\Psi}+
    \frac{2}{h}\right)h_{,\theta}^{2}
  +2\frac{\Psi_{,\theta}}{\Psi}
  h_{,\theta}\left(1+\frac{h_{,\theta}^{2}}{h^{2}}\right)=0.
\end{equation}
We solve this equation with the boundary condition $h_{,\theta}=0$ at
$\theta=0,\pi/2$.  The ADM mass in a conformally flat space is
defined as follows:
\begin{equation}
  M_{\rm{ADM}}=-\frac{1}{2\pi}\int_{r\rightarrow\infty}\partial_i\Psi 
N^idS~,
\end{equation}
where $N^i$ is a unit outward normal to a spherical surface at
infinity.  Using the Gauss law and the boundary condition at
$r^{(\pm)}=0$, we get
\begin{equation}
  M_{\rm{ADM}} = M_0+\frac{1}{2\pi}\int
  \frac18K^{ij}K_{ij}\frac{1}{\Psi^7}d\bs{r}'{}^3.
\end{equation}

We also calculate the minimum value $H^{(\rm{H})}$
of $H^{(\rm{H})}(\rm{S})$ with the Hawking mass.
The Hawking mass $M_{\rm{H}}(\rm{S})$ inside the surface S is defined
by
\begin{equation}
  M_{\rm{H}}(\text{S})=\frac{S^{1/2}}{(4\pi)^{3/2}}
\left( 2\pi+\frac{I(\rm{S})}{8}\right),
\end{equation}
where $I({\rm{S}})=\int_{{\rm{S}}}\theta_+\theta_-dS$ and 
$S$ is the area of the surface S.
$\theta_+$ and  $\theta_-$ are the expansion for the outgoing
 and the ingoing null congruence on
S, respectively. The minimum value $H^{\rm{(H)}}$ of $H^{(\rm{H})}(\rm{S})$
 is expressed as
\begin{equation}
H^{\rm{(H)}}=\frac{C}{4\pi M_H}=\frac{C}{\sqrt{\pi 
S}}\left(1+\frac{I}{16\pi}\right)^{-1}\label{CSI}
\end{equation}
where $C$ and $S$ are evaluated on the surface $\rm{S}_{\rm{min}}$ that gives the 
minimum value of $H^{(\rm{H})}(\rm{S})$ and $I\equiv I(\rm{S}_{\rm{min}})$.

We expand the equation of the 
surface $r=h(\theta)$ by using Legendre's polynormals:
\begin{equation}
  h(\theta)=a_0+a_2P_2(\cos\theta)+a_4P_4(\cos\theta )+\cdots .
\end{equation}
We truncate the series with some finite number $l_c$ and find the
coefficient $a_l~(l=0,2,\cdots,l_c)$ that make the value 
$H^{(\rm{H})}(\rm{S})$
minimum. We used $l_c=4$.  By using the different value $l_c=2$, 
we estimated that the maximal numerical
error is less than $0.1\%$.
%

\section{Numerical results}
\begin{figure}[hbtp]
\centering
\includegraphics[width=0.7\linewidth]{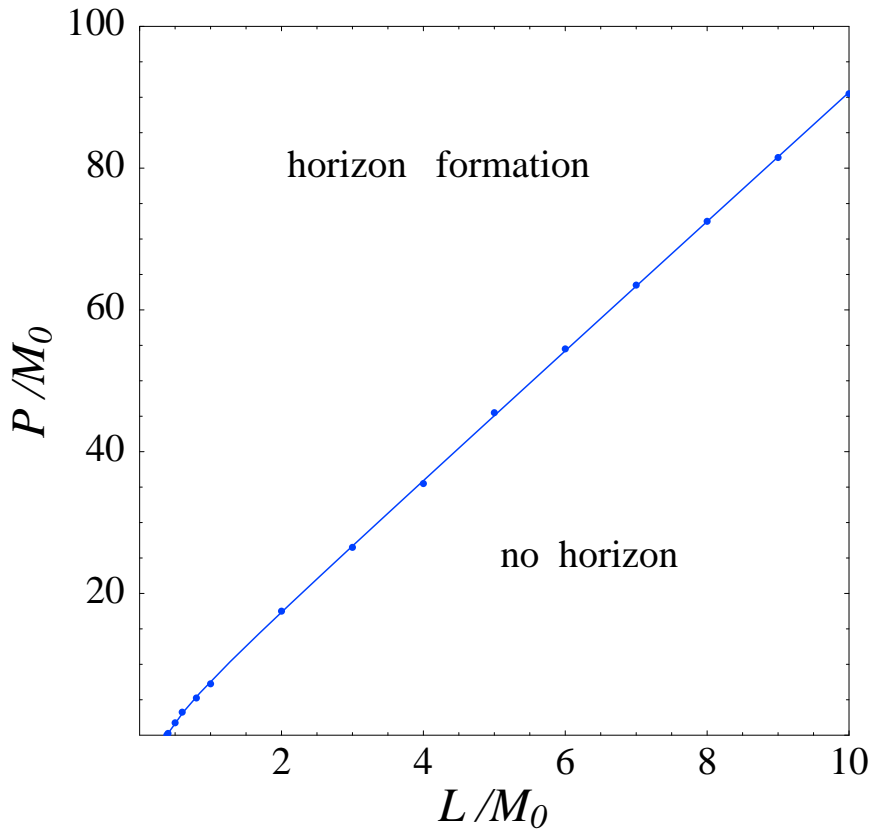}
\caption{
  The parameter space $(L,P)$ of the initial data. For the parameters in the
  upper region, an apparent horizon that encloses the two black holes
  exists. The momentum $P$ helps the formation of the horizon. Two regions 
are separated by the critical line.}
\label{fig:fig1}
\end{figure}

FIG.~\ref{fig:fig1} shows the parameter $(L,P)$ of the initial data
that the apparent horizon exists.
Even for large separation $L$ of  two black holes, an apparent horizon
can exist provided that the momentum $P$ is sufficiently large.
We call the boundary of these two regions as a
critical line $P=P_{\rm{crit}}(L)$.  The critical line intersects the
$L$-axis at $L=0.383M_0$.
The line $P=9.8L$ is an asymptote of the critical line
for $1 \ll L/M_0$ and $1 \ll P/M_0$.  Cook {\it et al.}~\cite{C} also
obtained a critical line for two colliding black holes in a similar
setting.  He used the Misner-Lindquist type initial data with two
Einstein-Rosen bridges and two asymptotically flat regions.  He also
got the result that the motion of two black holes helps the horizon
formation in the colliding case.

\begin{figure}[hbtp]
\centering
\includegraphics[width=0.7\linewidth]{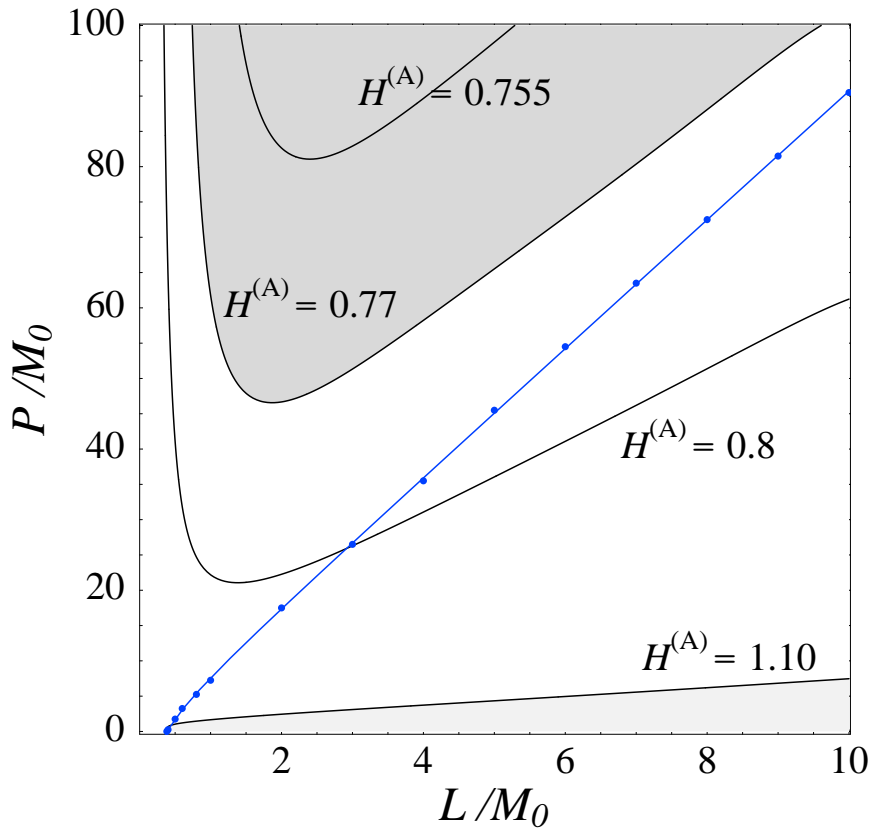}
\caption{
  The contour lines of $H^{(\rm{A})}=C/(4\pi M_{\text{ADM}})$.
   For the initial data with parameters in the light gray region ($1.10
  <H^{(\rm{A})}$), there is no apparent horizon. For the initial data with
  parameters in the dark gray
  region ($H^{(\rm{A})}<0.77$), an apparent horizon always
  exists. We have two values $H_1^{(\rm{A})}=0.77$ and
  $H_2^{(\rm{A})}=1.10$.}
\label{fig:fig2}
\end{figure}

Now we test the hoop conjecture with the ADM mass.  FIG.~\ref{fig:fig2} shows the
contour lines of $H^{(\rm{A})}$ in the $(L,P)$-plane.
%
%
The contour line with $H^{(\rm{A})}=1.10$ becomes a tangent line of
the critical line near $(L, P)\sim (0.4M_0,0)$.  We can
observe that the apparent horizon never exists for $1.10<H^{(\rm{A})}$
and we have $H^{(\rm{A})}_2=1.10$.
For large $L$ and $P$, the contour line with 
$H^{(\rm{A})}=0.77$
becomes an asymptote of the critical line.  For $H^{(\rm{A})}<0.77$,
a black hole horizon always exists and we have $H^{(A)}_1=0.77$.

\begin{figure}[hbtp]
\centering
\includegraphics[width=0.7\linewidth]{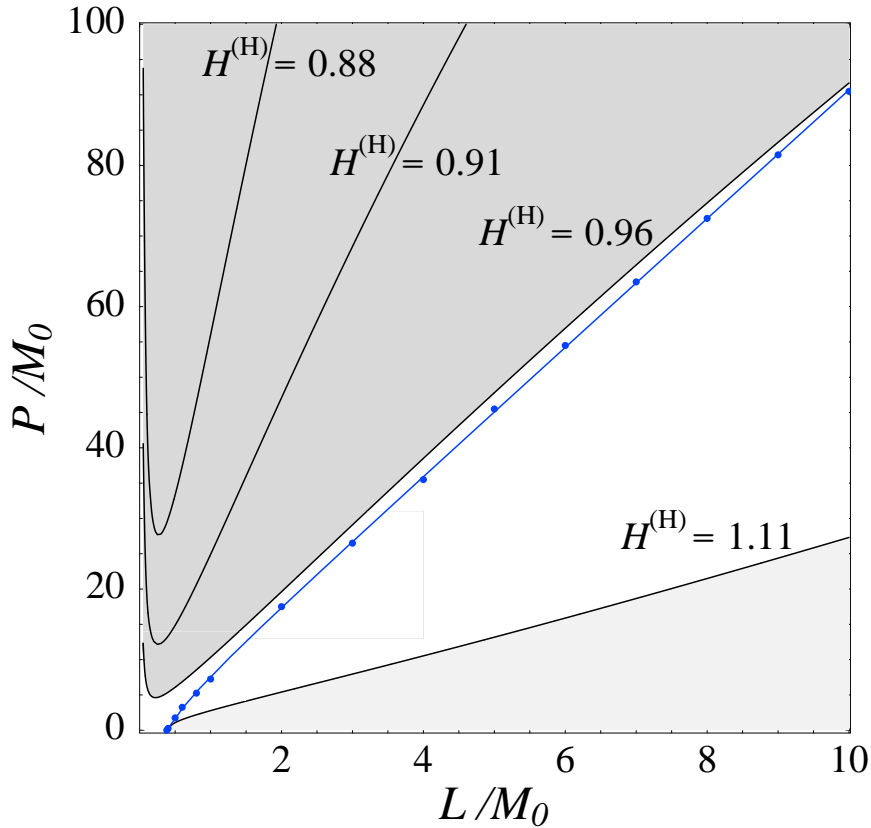}
\caption{
  The contour lines of $H^{(\rm{H})}=C/(4\pi M_{\text{H}})$.
   For the initial data with parameters in the light gray region (
  $1.11<H^{(\rm{H})}$), there is no horizon. For the initial data with
  parameters in the dark
  gray region ($H^{(\rm{H})}<1.11$), an apparent horizon
  always exists. We have two values $H_1^{(\rm{H})}=0.96$ and
  $H_2^{(\rm{H})}=1.11$.  }
\label{fig:fig3}
\end{figure}

Next, we calculate the value $H^{(\rm{H})}$ using  the Hawking's quasi-local 
mass. The result is shown in FIG.~\ref{fig:fig3}.
The contour line with $H^{(\rm{H})}=1.11$ is a
tangent line of the critical
line near $(L, P)\sim (0.4M_0, 0)$. The 
apparent horizon never exists for $1.11<H^{(\rm{H})}$ and we have
$H^{(\rm{H})}_2=1.11$.  For large $L$ and $P$, the
contour line with $H^{(\rm{H})}=0.96$ becomes an asymptote of the
critical line.  For $H^{(\rm{H})}<0.96$,
a horizon always exists and we have $H^{(\rm{H})}_1=0.96$.

In both cases that we use the ADM mass and the Hawking mass, we
 obtain two values $H_1$ and $H_2$,  and the ratio 
$H$ is related to the existence of the horizon.
 As the kinetic energy due to the 
colliding
motion of black holes increases the rest mass of the system, the
increase of the momentum $P$ results in the increase of the mass.  This
leads to the decrease of $H$ in accordance with the horizon formation.
In the $(L,P)$-plane, the region $H^{(\rm{H})}<H^{(\rm{H})}_1$ contains
the region $H^{(\rm{A})}<H^{(\rm{A})}_1$, and
the region $H^{(\rm{H})}_2<H^{(\rm{H})}$ contains
the region $H^{(\rm{A})}_2<H^{(\rm{A})}$. Hence $H^{(\rm{H})}$ is
superior to $H^{(\rm{A})}$ because there are some cases that we can
judge the existence of a horizon
by using $H^{(\rm{H})}$ but cannot judge
it by using $H^{(\rm{A})}$.

We explain the reason why $H^{(\rm{H})}$ is the better parameter
for the horizon formation compared to $H^{(\rm{A})}$ 
as follows. The Bowen-York initial data for two black holes contains a lot 
of gravitational wave mode which is not localized so much. 
The ADM mass evaluates 
not only the gravitational energy that is the source to
make a black hole horizon, but also the gravitational energy 
that does not contributes to the horizon formation.  As the Hawking mass 
gives proper
gravitational mass, $H^{(\rm{H})}$
gives a better picture of the hoop conjecture compared to $H^{(\rm{A})}$.

\begin{figure}[hbtp]
\centering
\includegraphics[width=0.7\linewidth]{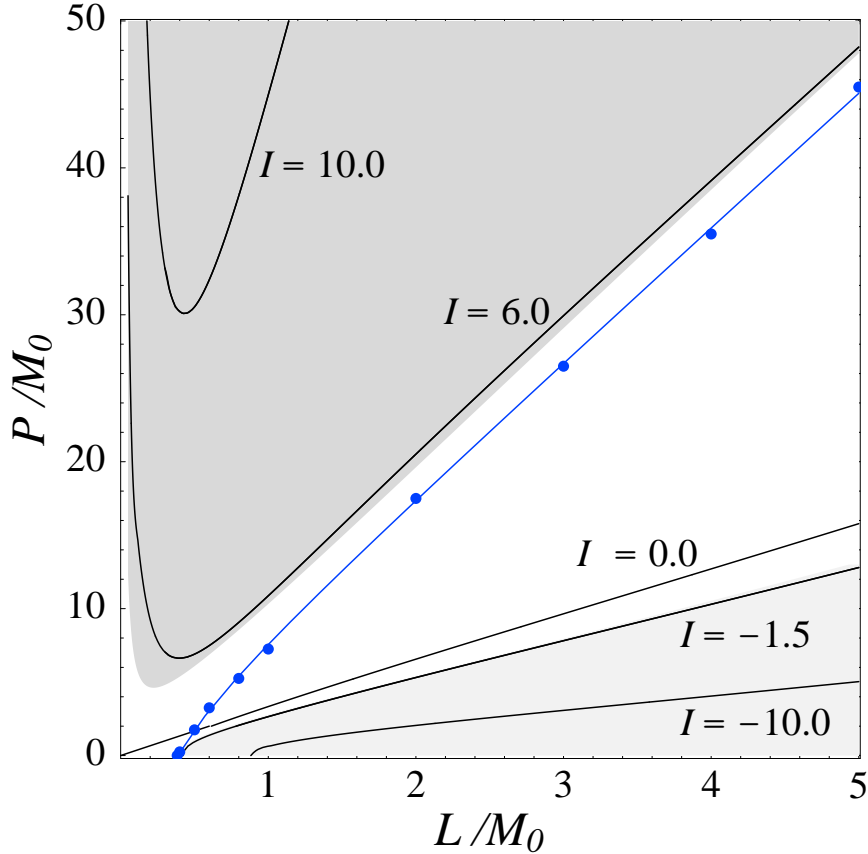}
\caption{
  The contour lines of the integral $I=\int_{\text{S}}\theta_+\theta_-dS$ in 
the Hawking mass.
  The regions $H^{(\textrm{H})}_2<H^{(\textrm{H})}$ (light grey region) and 
$H^{(\textrm{H})}<H^{(\textrm{H})}_1$ (dark grey region) are also shown.
     }
\label{fig:fig4}
\end{figure}

In introduction, we stated that
$H^{\rm{(H)}}$ has a critical value for the horizon formation
 in the spherically symmetric case and it is the desirable feature of
$H^{(\rm{H})}$.
For the spherically symmetric spaces, 
$C/\sqrt{\pi S}=1$ and Eq.~\eqref{CSI} becomes 
$H^{(\rm{H})}=(1+I/16\pi)^{-1}$.
If we restrict our attention to the spherically symmetric space without 
white hole regions, $\theta_-$ is negative for all surfaces S.
Because $I(\textrm{S})=\int_{\textrm{S}}\theta_+\theta_-dS=\theta_+\theta_-S$,
the sign of $I(\rm{S})$ is opposite to the sign of $\theta_+$ on S.
If there is no apparent horizon, $\theta_+>0$ and $I(\rm{S})<0$ for all
surfaces S. Thus $I$ is negative and $H^{(\rm{H})}$ becomes greater than unity.
 If there is an apparent horizon, there exists a surface S on which 
 $\theta_+\le 0$  and $I(\rm{S})\ge 0$.
Thus $I\ge 0$ and $H^{(\rm{H})}$ becomes smaller than unity. 
Therefore in the spherically symmetric cases, the sign of $I$ exactly 
corresponds to the existence of an
apparent horizon and $H^{(\rm{H})}$ becomes a critical parameter for 
the horizon formation.

In the two black hole case, the factor $C/\sqrt{\pi S}$ in Eq.~\eqref{CSI}
varies in the range $1.0\leq C/\sqrt{\pi S} \leq 1.1$ for the parameters 
$(L,P)$ that we
have investigated. Hence the value of $H^{\rm{(H)}}$ depends mainly on the value 
of $I$.
The contour lines of $I$ in the $(L,P)$-plane is shown in FIG.~\ref{fig:fig4}.
The  line  with $I=0$ that can be approximated as $P=3.2L$
crosses the critical line. In contrast to the case of the 
spherical collapse, the sign of $I$ does not exactly
corresponds to the existence of a horizon.
There exist parameter regions  that $I<0$ with a 
horizon and $I>0$ without a horizon.
In $P=0$ case,  
$\theta_+=-\theta_-$ holds on S and $I$ always becomes negative, even there
is a horizon. Because $I$ depends continuously on parameters $L$ and $P$, 
there is a region $P_{\rm{crit}}(L)<P<3.2L$ where $I<0$
with a horizon.
By the equation \eqref{CSI}, this negative $I$ implies that
$H^{\rm{(H)}}>{C}/{\sqrt{\pi S}}>1$.
Thus there is a region in $(L,P)$-plane
where $H^{\rm{(H)}}$ takes the value greater than $1$ but there is a 
horizon.
For the initial data with parameters $3.2L<P<P_{\rm{crit}}(L)$,
a horizon does not exist and there is no surface
on which $\theta_+\theta_-$ keeps positive definite value. But the sign of 
the integral $I$ is positive.
 This is because for non-time-symmetric initial data, 
it is possible to make a surface on which $\theta_+\theta_-$ changes its sign. 
Thus there exists a surface S that satisfy $I(\text{S})>0$
even there is no horizon.
On the surface $\rm{S}_{\rm{min}}$,
$\theta_+\theta_-$ is positive near the poles and this makes the integral
$I$ positive. In the region
$3.2L<P<P_{\rm{crit}}(L)$, $I$ takes a value up to $6.0$ . 
For $I\simeq 6$, $H^{(\rm{H})} \simeq 0.89C/\sqrt{\pi S}<1$.
This is the reason why there is the parameter region that 
 $H^{(\text{H})}$ takes the value less than unity
but the apparent horizon does not exist.

In the two black hole case, the sign of $I$ does not exactly correspond
to the existence of a horizon.
A horizon exists if $I>6$, and a horizon does not exist if $I<-1.5$.
These correspond to the values $H^{\rm{(H)}}_1=0.96$ and $H^{\rm{(H)}}_2=1.11$.
Contrary to the spherically symmetric case, 
 $H^{(\rm{H})}$ does not have a critical value
for the horizon formation in the axially symmetric case.
But $H^{(\rm{H})}$ is the better parameter than  $H^{(\rm{A})}$ to judge the horizon formation.

\section{Summary and discussion}

In the context of the hoop conjecture, we investigated the condition
for the apparent horizon formation for two colliding black holes. The
motion of black holes helps the formation of the horizon.  We
tested the hoop conjecture using the ADM mass and the Hawking mass.
Although in both cases the ratio $H$ is related to the existence of
the horizon, $H^{(\rm{H})}$ is superior to $H^{(\rm{A})}$ for the
purpose of judging the existence of the horizon.


In this paper, we considered the  two black hole system only in colliding
case.  We have also investigated the receding case $P<0$.  When two
black holes are receding, the motion prevents the formation of the
black hole horizon. But with the increase of $|P|$, the value $H$
decreases and we have no value $H_1$.  This behavior of $H$ shows that
the ratio $H$ is not a parameter for the black hole formation in
the receding case.  However in the receding case, $H$ is related to
the formation of a white hole horizon on which $\theta_-=0$.
Combining the colliding and the receding cases, we can say that
$H$ is a parameter of either the black hole formation or
the white hole formation.

In the receding case, if a black hole horizon exists,
it is located in a white hole region.  Even S is in a black hole region,
$I(\rm{S})$ becomes negative because $\theta_+<0$ and
$\theta_->0$ are satisfied on S.
Thus the sign of $I$ scarcely corresponds to the existence of the black hole 
horizon.
This is the reason why $H^{(\rm{H})}$ cannot be a parameter
for the black hole formation in the receding case. Similarly,
in spherically symmetric case, $H^{(\rm{H})}$ cannot be a
parameter for the black hole horizon formation if white hole horizon
exists outside of it.

The formation of a
black hole  occurs as the result of the temporal evolution of the
initial data, and  a black hole horizon in a white hole region may not
be realized by the process of usual gravitational collapse if the initial 
data does not contain white hole regions.
As far as the physically realizable situation is concerned,
we expect that $H^{(\rm{H})}$ is
a good parameter to judge  the black hole formation.


\end{document}